\documentstyle[11pt,amssymb,epsf]{article}

\def\ben{\begin{equation}}
\def\een{\end{equation}}

\let\a=\alpha    
    
  \let\n=\nu

\let\C=\Chi

\def\nn{\nonumber} \def\bd{\begin{document}} \def\ed{\end{document}}
\def\ds{\documentstyle} \let\fr=\frac \let\bl=\bigl \let\br=\bigr
\let\Br=\Bigr \let\Bl=\Bigl
\let\bm=\bibitem
\let\na=\nabla
\let\pa=\partial \let\ov=\overline
\newcommand{\be}{\begin{equation}}
\newcommand{\ee}{\end{equation}}
\def\ba{\begin{array}}
\def\ea{\end{array}}
\def\ft#1#2{{\textstyle{{\scriptstyle #1}\over {\scriptstyle #2}}}}
\def\fft#1#2{{#1 \over #2}}
\def\del{\partial}
\def\vp{\varphi}
\def\sst#1{{\scriptscriptstyle #1}}
\def\oneone{\rlap 1\mkern4mu{\rm l}}
\def\td{\tilde}
\def\wtd{\widetilde}
\def\ie{\rm i.e.\ }
\def\dalemb#1#2{{\vbox{\hrule height .#2pt
        \hbox{\vrule width.#2pt height#1pt \kern#1pt
                \vrule width.#2pt}
        \hrule height.#2pt}}}
\def\square{\mathord{\dalemb{6.8}{7}\hbox{\hskip1pt}}}
\newcommand{\ho}[1]{$\, ^{#1}$}
\newcommand{\hoch}[1]{$\, ^{#1}$}
\newcommand{\bea}{\begin{eqnarray}}
\newcommand{\eea}{\end{eqnarray}}
\newcommand{\ra}{\rightarrow}
\newcommand{\lra}{\longrightarrow}
\newcommand{\Lra}{\Leftrightarrow}
\newcommand{\bp}{\tilde \beta^\prime}
\newcommand{\tr}{{\rm tr} }
\newcommand{\Tr}{{\rm Tr} }
\def\0{{\sst{(0)}}}
\def\1{{\sst{(1)}}}
\def\2{{\sst{(2)}}}
\def\3{{\sst{(3)}}}
\def\4{{\sst{(4)}}}
\def\5{{\sst{(5)}}}
\def\6{{\sst{(6)}}}
\def\7{{\sst{(7)}}}
\def\8{{\sst{(8)}}}
\def\n{{\sst{(n)}}}
\def\cA{{{\cal A}}}
\def\cB{{{\cal B}}}
\def\cF{{{\cal F}}}
\def\cH{{{\cal H}}}
\def\tV{\widetilde V}
\def\tW{\widetilde W}
\def\tH{\widetilde H}
\def\tE{\widetilde E}
\def\tF{\widetilde F}
\def\tA{\widetilde A}
\def\im{{i}}
\def\tY{{{\wtd Y}}}
\def\ep{{\epsilon}}
\def\vep{{\varepsilon}}
\def\R{\rlap{\rm I}\mkern3mu{\rm R}}
\def\bD{{{\bar D}}}

\def\R{\rlap{\rm I}\mkern3mu{\rm R}}
\def\bD{{{\bar D}}}
\def\R{{{\Bbb R}}}
\def\C{{{\Bbb C}}}
\def\H{{{\Bbb H}}}
\def\CP{{{\Bbb C}{\Bbb P}}}
\def\RP{{{\Bbb R}{\Bbb P}}}
\def\Z{{{\Bbb Z}}}
\def\bA{{{\Bbb A}}}
\def\bB{{{\Bbb B}}}
\def\bC{{{\Bbb C}}}
\def\bD{{{\Bbb D}}}
\def\bE{{{\Bbb E}}}
\def\bZ{{{\Bbb Z}}}
\def\Re{{{\frak{Re}}}}
\def\Im{{{\frak{Im}}}}
\def\cosec{{\,\hbox{cosec}\,}}
\def\Gm{{\Gamma_{\!\! -}}}
\def\Gp{{\Gamma_{\!\! +}}}
\def\stan{{standard }}
\def\nonstan{{supernumerary }}

\newcommand{\tamphys}{\it Center for Theoretical Physics,
Texas A\&M University, College Station, TX 77843}

\newcommand{\upenn}{\it Department of Physics and Astronomy,\\ University
of Pennsylvania, Philadelphia, PA 19104}

\newcommand{\brussels}{\it Physique Th\'eorique et Math\'ematique,
Universit\'e Libre de Bruxelles,\\ Campus Plaine C.P. 231, B-1050
Bruxelles, Belgium}

\newcommand{\auth}{Z.-W. Chong\hoch{\ddagger1},  
H. L\"u\hoch{\ddagger1} and C.N. Pope\hoch{\ddagger1}}

\begin{document}

\begin{flushright}
MIFP-04-26 \\
{\bf hep-th/0412221}\\
December\  2004
\end{flushright}

\vspace{10pt}

\begin{center}

{\Large {\bf BPS Geometries and AdS Bubbles }}

\vspace{20pt}
\auth

\vspace{10pt}{\hoch{\dagger}\it George P. \& Cynthia W. Mitchell
Institute for Fundamental Physics,\\ Texas A\& M University,
College Station, TX 77843-4242, USA}


%
%
%

\vspace{20pt}


\begin{abstract}

    Recently, $\ft12$-BPS AdS bubble solutions have been obtained by
Lin, Lunin and Maldacena, which correspond to Fermi droplets in phase
space in the dual CFT picture.  They can be thought of as
generalisations of $\ft12$-BPS AdS black hole solutions in five or
seven dimensional gauged supergravity.  In this paper, we extend these
solutions by invoking additional gauge fields and scalar fields in the
supergravity Lagrangians, thereby obtaining AdS bubble generalisations
of the previously-known multi-charge AdS black solutions of gauged
supergravity.  We also obtain analogous AdS bubble solutions in
four-dimensional gauged supergravity.  Our solutions generically
preserve supersymmetry fractions $\ft14$, $\ft18$ and $\ft18$ in
seven, five and four dimensions respectively.  They can be lifted to
M-theory or type IIB string theory, using previously known formulae
for the consistent Pauli sphere reductions that yield the gauged
supergravities.  We also find similar solutions in six-dimensional 
gauged supergravity, and discuss their lift to the massive type IIA
theory.

\end{abstract}
\end{center}

{\vfill\leftline{}\vfill \vskip 10pt \footnoterule {\footnotesize
{\footnotesize
\hoch{1} Research supported in part by DOE grant
DE-FG03-95ER40917.}\vskip 2pt
}

\pagebreak

\newpage

\section{Introduction}

   It was recently shown in \cite{linlunmal} that there exist smooth
BPS solutions in the ten-dimensional type IIB string and in M-theory,
which preserve one half of the supersymmetry, and which admit an
interpretation in the dual field theory, via the AdS/CFT
correspondence, as droplets in a phase space occupied by fermions.  In
\cite{linlunmal}, the $\ft12$-BPS M-theory solutions were obtained by
first constructing solutions in seven-dimensional gauged supergravity,
and then lifting them to eleven dimensions by making use of the
consistent $S^4$ reduction that was constructed in \cite{vann1,vann2}.
The seven-dimensional solutions can be thought of as certain
generalisations of $\ft12$-BPS black-hole solutions of
seven-dimensional gauged supergravity, which are contained within a
construction of seven-dimensional AdS black holes in \cite{cvetgubs}.
The generalisation involves turning on an additional scalar field that
is set to zero in the original black-hole solution.  Interestingly,
the resulting ``AdS bubble'' solutions obtained in \cite{linlunmal}
are in general everywhere smooth, unlike the $\ft12$-BPS extremal AdS
black hole which suffers from a naked singularity that lies strictly
outside the horizon.

   The $\ft12$-BPS solutions of type IIB supergravity that were obtained
in \cite{linlunmal} were found by a direct ten-dimensional 
construction.\footnote{Analogous solutions in $D=6$ were found recently
in \cite{liu}.}
They can also be viewed from a five-dimensional point of view, by
making use of results obtained in \cite{cvluposatr5} on the consistent
reduction of type IIB supergravity on $S^5$.  The type IIB solutions in
\cite{linlunmal} then acquire an interpretation as generalisations of
$\ft12$-BPS AdS black holes in five-dimensional supergravity, which were
contained within a construction of five-dimensional AdS black holes 
in \cite{behcvesab}.

   Since the AdS bubble solutions obtained in \cite{linlunmal} are
generalisations of $\ft12$-supersymmetric AdS black holes in $D=7$ and
$D=5$, which carry just a single electric charge, a further extension
naturally suggests itself, in which one turns on additional electric
charges, thereby giving solutions which preserve smaller supersymmetry
fractions.  For the AdS black hole solutions themselves, one can turn
on two independent charges in $D=7$, giving solutions which preserve
$\ft14$ supersymmetry in the BPS limit.  Likewise, in five dimensions
one can obtain AdS black hole solutions with three independent
charges, which have BPS limits preserving $\ft18$ of the supersymmetry when
all three charges are non-zero.

   In this paper, we construct ``AdS bubble'' solutions in $D=7$,
$D=5$ and $D=4$ gauged supergravities, which respectively generalise
the 2-charge, 3-charge and 4-charge BPS black hole solutions of these
theories.  Making use of results in \cite{vann1,vann2} for $D=7$, in 
\cite{cvluposatr5} for $D=5$, and in \cite{cvluposa,cvlupoconsist,tenauth},
we can straightforwardly lift all of the resulting solutions to ten or
eleven dimensions.  We therefore find AdS bubble solutions
that generically preserve $\ft14$ of the supersymmetry in seven dimensions, 
$\ft18$ of the supersymmetry in five dimensions, and $\ft18$ of the 
supersymmetry in four dimensions.  We also obtain AdS bubble solutions
in six-dimensional gauged supergravity, and discuss their lift to
the massive type IIA theory.

\section{$\ft14$-BPS Geometries in Seven Dimensions}\label{d7sec}

   The relevant part of the ${\cal N}=4$ gauged supergravity theory
that suffices for describing the supersymmetric solutions with $\ft12$
and $\ft14$ supersymmetry is derivable from the
Lagrangian\footnote{This corresponds to a consistent truncation of the
full $SO(5)$ gauged supergravity, provided that one restricts the gauge
fields to satisfy $F_\2^{[ij}\wedge F_\2^{k\ell]}=0$, so that
the 3-form fields are not excited.  This restriction is indeed satisfied
in the AdS bubble solutions we shall consider.}
\be
{\cal L} = R\, {*\oneone} - \ft14 T_{ij}^{-1}\, {*D T_{jk}}\wedge 
   T_{k\ell}^{-1}\, D T_{\ell i} - \ft14 T_{ik}^{-1}\, 
     T_{j\ell}^{-1}\, {*F_\2^{ij}}\wedge F_\2^{k\ell} - V\, {*\oneone}\,,
\label{genlag}
\ee
where the 14 scalars are described by the unimodular symmetric matrix
$T_{ij}$, the covariant derivative is defined by
\be
D T_{ij} = dT_{ij} + g\, A_\1^{ik}\, T_{kj} + g\, A_\1^{jk}\, T_{ik}\,,
\ee
and the $SO(5)$ gauge fields are given by $F_\2^{ij} = dA_\1^{ij} +
 g A_\1^{ik}\wedge A_\1^{kj}$.  The scalar potential $V$ is given by
\be
V= \ft12 g^2\, (2T_{ij}\, T_{ij} - T_{ii}^2)\,.\label{VT}
\ee

   We consider the following restriction of the gauge and scalar fields, 
which arises as a consistent truncation of the complete set of
equations of motion:
\bea
A_\1^{12} &=& A_\1^1\,,\qquad A_\1^{34} = A_\1^2\,,\nn\\
T_{ij} &=& \hbox{\rm{diag}}(X_1\, e^{-\vp_1}, X_1\, e^{\vp_1}, 
          X_2\, e^{-\vp_2}, X_2\, e^{\vp_2}, (X_1\, X_2)^{-2})\,,
\eea
with all other gauge fields being zero.  This corresponds to a truncation
to gauge fields in the maximal torus $U(1)\times U(1)$ in $SO(5)$.  We are
retaining four scalar fields in the truncation, namely $(X_1,X_2, \vp_1,
\vp_2)$.   In terms of a canonical parameterisation, we can write
\be
X_1 = e^{-\ft1{\sqrt2}\, \phi_1 - \ft1{\sqrt{10}}\, \phi_2}\,\qquad
  X_2= e^{\ft1{\sqrt2}\, \phi_1 - \ft1{\sqrt{10}}\, \phi_2}\,,
\ee
whereupon the Lagrangian for this truncated system becomes
\bea
{\cal L}_7 &=& R\, {*\oneone} - \ft12 \sum_{i=1}^2\, 
   {*d\vp_i}\wedge d\vp_i -\ft12 \sum_{\a=1}^2 {*d\phi_\a}\wedge d\phi_\a 
- \ft12 \sum_{i=1}^2 X_i^{-2}\, {*F_\2^i}\wedge 
        F_\2^i \nn\\
&&-2 g^2\, 
   \sum_{i=1}^2 \sinh^2\vp_i\, {*A_\1^i}\wedge A_\1^i- V\, {*\oneone}\,,
\label{d7trunc}
\eea
with the potential's being given by
\bea
V &=& g^2\, ( 2X_1^2\, \sinh^2 \vp_1 + 2 X_2^2\, \sinh^2\vp_2 -
    2 X_1^{-1}\, X_2^{-2}\, \cosh\vp_1 - 2 X_2^{-1}\, X_1^{-2}\, 
   \cosh\vp_2 \nn\\
&&\qquad - 4 X_1 X_2\, \cosh\vp_1\, \cosh\vp_2  + 
    \ft12 X_1^{-4}\, X_2^{-4})\,.
\eea

   Guided by the form of the previously-known 2-charge black-hole 
solutions \cite{cvetgubs,tenauth},
and the recent single-charge ``AdS bubble'' solutions obtained in
\cite{linlunmal}, we are led to propose the following ansatz for general 
2-charge seven-dimensional AdS bubble solutions:
\bea
ds_7^2 &=& -(H_1 H_2)^{-4/5}\, f\, dt^2 + (H_1 H_2)^{1/5}\, (f^{-1}\, 
   dr^2 + r^2 \, d\Omega_5^2)\,,\nn\\
A_\1^i &=& - H_i^{-1}\, dt\,,\qquad X_i= (H_1 H_2)^{2/5}\, H_i^{-1}\,,
\qquad   \cosh\vp_i= (R H_i)'\,,\nn\\
f&=& 1 + \ft14 g^2\, r^2\, H_1 H_2\,,\label{d7ans}
\eea
where $R\equiv \ft14 r^4$ and a prime denotes a derivative with respect
to $R$. 

   Substituting into the equations of motion that follow from 
(\ref{d7trunc}), we find that they are indeed satisfied by the ansatz 
(\ref{d7ans}), provided that the functions $H_i$ satisfy the equations
\be
2 R^{1/2}\, f\, (R H_i)'' = - g^2\, [{(R H_i)'}^2- 1](H_1 H_2)\, H_i^{-1}\,.
\ee
The 2-charge solutions that we have obtained generalise the single-charge
$\ft12$-BPS solutions of \cite{linlunmal}.  When both charges are non-zero,
our solutions preserve $\ft14$ of the supersymmetry.  If one or other of
the gauge fields is set to zero (achieved by setting $H_1=1$ or $H_2=1$),
the solutions reduce to those in \cite{linlunmal}.

\section{$\ft18$-BPS Geometries in Five Dimensions}\label{d5sec}

   In a similar fashion, we can construct 3-charge AdS bubble
solutions in five dimensions that correspond to generalisations of the
$\ft12$-BPS solutions of type IIB supergravity that were obtained in
\cite{linlunmal}.  Although they were not described in gauged
five-dimensional supergravity in \cite{linlunmal}, the type IIB
solutions obtained there can be viewed as the lifting, via the
consistent $S^5$ Pauli reduction obtained in \cite{cvluposatr5}, of
single-charge solutions in five-dimensional gauged supergravity.  Our
generalisation is obtained by considering a consistent truncation of
five-dimensional maximal gauged supergravity in which gauge fields in
the $U(1)^3$ maximal torus of the $SO(6)$ gauge group are retained,
together with a total of five scalar fields.  Prior to the truncation
of the $SO(6)$ gauge group, a consistently truncated subset of the
bosonic fields, comprising the $SO(6)$ gauge fields and an irreducible
20 of scalars, together with the metric, can be described by the
Lagrangian (\ref{genlag}), where now the $i, j$ indices are in the
vector representation of $SO(6)$.  The further consistent
truncation that we are making then involves setting
\be
A_\1^{12}= A_\1^1\,,\qquad A_\1^{34} = A_\1^2\,,\qquad A_\1^{56}= A_\1^3\,,
\ee
with all other gauge fields (except for those implied by antisymmetry in
$ij$) set to zero.  The scalar fields that we retain are given by
\be
T_{ij}= \hbox{\rm diag}(X_1\, e^{-\vp_1}, X_1\, e^{\vp_1}, 
                     X_2\, e^{-\vp_2}, X_2\, e^{\vp_2}, 
                    X_3\, e^{-\vp_3}, X_3\, e^{\vp_3})\,.\label{5scal}
\ee
The Lagrangian that describes this truncated system is given by
\bea
{\cal L}_5 &=& R\, {*\oneone} - \ft12 \sum_{i=1}^3\, 
   {*d\vp_i}\wedge d\vp_i 
  -\ft12 \sum_{\a=1}^2 {*d\phi_\a}\wedge d\phi_\a 
 - \ft12 \sum_{i=1}^3 X_i^{-2}\, {*F_\2^i}\wedge 
        F_\2^i \nn\\
&&-2 g^2\, 
   \sum_{i=1}^3 \sinh^2\vp_i\, {*A_\1^i}\wedge A_\1^i- V\, {*\oneone}
   + F_\2^1\wedge F_\2^2 \wedge A_\1^3 \,,
\label{d5trunc}
\eea
where we write
\be
X_1 = e^{-\ft1{\sqrt6}\, \phi_1 -\ft1{\sqrt2}\, \phi_2}\,,\qquad
X_2 = e^{-\ft1{\sqrt6}\, \phi_1 +\ft1{\sqrt2}\, \phi_2}\,,\qquad
X_3= e^{\ft2{\sqrt{6}}\, \phi_2}\,.
\ee
Note that $X_1 X_2 X_3=1$.  The scalar potential following from
(\ref{VT}) and (\ref{5scal}) is
\bea
V &=& 2g^2\, (X_1^2\, \sinh^2\vp_1 +  X_2^2\, \sinh^2\vp_2 +
              X_3^2\, \sinh^2\vp_3 - 2 X_2 X_3 \cosh\vp_2\, \cosh\vp_3 
\nn\\
&& \qquad - 2 X_3 X_1 \cosh\vp_3\, \cosh\vp_1
                - 2 X_1 X_2 \cosh\vp_1\, \cosh\vp_2)\,.
\eea

   Based on the previously-known black-hole solutions in
five-dimensional gauged supergravity \cite{behcvesab}, and the
seven-dimensional single-charge AdS bubble metrics of \cite{linlunmal}
and their 2-charge generalisation that we found in section
\ref{d7sec}, we are led to the following ansatz for 3-charge AdS
bubble solutions of five-dimensional gauged supergravity:
\bea
ds_5^2 &=& -(H_1 H_2 H_3)^{-2/3}\, f\, dt^2 + (H_1 H_2 H_3)^{1/3}\, (f^{-1}\, 
   dr^2 + r^2 \, d\Omega_3^2)\,,\nn\\
A_\1^i &=& - H_i^{-1}\, dt\,,\qquad X_i= (H_1 H_2 H_3 )^{1/3}\, H_i^{-1}\,,
\qquad  \cosh\vp_i= (R H_i)'\,,\nn\\
f&=& 1 +  g^2\, r^2\, H_1 H_2 H_3\,,\label{d5ans}
\eea
where $R\equiv r^2$ and a prime denotes a derivative with respect
to $R$.  
Substituting into the equations of motion following from 
(\ref{d5trunc}), we find that they are indeed solutions, provided that
the functions $H_i$ obey the equations
\be
f\, (R H_i)'' = - g^2\, [{(R H_i)'}^2- 1]\, (H_1 H_2 H_3)\, H_i^{-1}\,,
\ee
where the prime denotes the derivative with respect to $R$.

  These solutions in general preserve $\ft18$ of the maximal
supersymmetry in five dimensions.  In general, if only $N$ of the
maximum of 3 charges is turned on, the solution preserves a fraction
$2^{-N}$ of the supersymmetry.  The solutions can be lifted to give
solutions of the type IIB theory in ten dimensions, using the formulae
obtained in \cite{cvluposatr5}.  If, in particular, only one of the
three gauge fields is taken to be non-zero (achieved by setting two of
the three $H_i$ functions to 1), the resulting $\ft12$-supersymmetric
configurations will coincide with solutions obtained in
\cite{linlunmal}.

\section{$\ft18$-BPS Geometries in Four Dimensions}

   The discussion of the previous two sections can be extended also to
$SO(8)$ gauged supergravity in four dimensions.  In this case, we can 
construct AdS bubble solutions that generalise the 4-charge BPS 
solutions that are contained within the AdS black hole solutions constructed 
in \cite{dufliu}.

   The relevant truncation of the full four-dimensional $SO(8)$ gauged
supergravity, containing the fields utilised in the AdS bubble
solutions, is described again by the Lagrangian (\ref{genlag}), where
now the indices $i$, $j$ lie in the vector representation of $SO(8)$.
Since the reduction ansatz for obtaining this truncation via a reduction 
from $D=11$ on $S^7$ has not appeared
explicitly in the literature in a coherent form, here we draw together
various presentations of parts of the ansatz, including results in
\cite{cvluposa,cvlupoconsist,tenauth}, to give the full expressions.
The reduction ansatz for the eleven-dimensional metric is
\be
d\hat s_{11}^2 = \Delta^{2/3}\, ds_4^2 + g^{-2}\, \Delta^{-1/3}\, 
     T_{ij}^{-1}\, D\mu^i\, D\mu^j\,,
\ee
where $\mu^i\, \mu^i=1$ defines the unit 7-sphere, and 
\be
\Delta\equiv T_{ij}\, \mu^i\, \mu^j\,,\qquad  
D\mu^i \equiv 
d\mu^i + g\, A^{ij}_\1\, \mu^j\,.
\ee
The 4-form field strength is reduced according to the ansatz
\be
\hat F_\4 = - g\, U\, \ep_\4 + g^{-1}\, T_{ij}^{-1}\, {*D T_{jk}}\wedge
   (\mu^k\, D\mu^i) - \ft12 g^{-2}\, T_{ik}^{-1}\, T_{j\ell}^{-1}\, 
{*F_\2^{ij}}\wedge D\mu^k\wedge D\mu^\ell\,,
\ee
where $U\equiv 2 T_{ij}\, T_{jk}\, \mu^i\, \mu^k - \Delta\, T_{ii}$,
and $\ep_\4$ is the volume form in the lower-dimensional metric
$ds_4^2$.  It should be noted that this reduction is fully consistent
{\it provided} that one restricts the field-strengths to
configurations for which $F_\2^{[ij}\wedge F_\2^{k\ell]}=0$, as indeed
we shall be doing in the AdS bubble solutions obtained below, which
carry only electric charges.  For generic configurations where the
quadratic products of field strengths are non-vanishing, these products
would be acting as sources for the 35 additional pseudo-scalar fields 
of ${\cal N}=8$ supergravity, and the reduction ansatz would be immeasurably
more complicated.  Fortunately, this complication need not concern us here.

    We then make a further consistent truncation that retains just the
metric, the four gauge fields of the $U(1)^4$ maximal torus in
$SO(8)$, and seven scalar fields:
\bea
A_\1^{12} &=& A_\1^1\,,\qquad 
A_\1^{34} = A_\1^2\,,\qquad
A_\1^{56} = A_\1^3\,,\qquad 
A_\1^{78} = A_\1^4\,, \\
T_{ij}&=&\hbox{\rm diag}(X_1\, e^{-\varphi_1},X_1\, e^{\varphi_1},
      X_2\, e^{-\varphi_2},X_2\, e^{\varphi_2},
      X_3\, e^{-\varphi_3},X_3\, e^{\varphi_3},
      X_4\, e^{-\varphi_4},X_4\, e^{\varphi_4})\,.\nn
\eea
The scalars $X_i$, which satisfy $X_1 X_2 X_3 X_4=1$, can be parameterised
canonically in the form
\be
X_1 = e^{\ft12 (-\phi_1 - \phi_2 - \phi_3)}\,,\quad
X_2 = e^{\ft12 (-\phi_1 + \phi_2 + \phi_3)}\,,\quad
X_3 = e^{\ft12 (\phi_1 - \phi_2 + \phi_3)}\,,\quad
X_4 = e^{\ft12 (\phi_1 + \phi_2 - \phi_3)}\,.
\ee

   Substituting the above into the expression (\ref{VT}) for the scalar
potential, we find that for this truncated system it is given by
\be
V= 2g^2\, \sum_{i=1}^4 X_i^2\, \sinh^2 \varphi_i - 2 g^2
   \sum_{i\ne j} X_i\, X_j \, \cosh\varphi_i\, \cosh\varphi_j\,.
\ee
The Lagrangian for the truncated system is given by
\bea
{\cal L}_4 &=& R\, {*\oneone} - \ft12 \sum_{i=1}^4\, 
   {*d\vp_i}\wedge d\vp_i 
-\ft12 \sum_{\a=1}^3 {*d\phi_\a}\wedge d\phi_\a
- \ft12 \sum_{i=1}^4 X_i^{-2}\, {*F_\2^i}\wedge 
        F_\2^i \nn\\
&& -2 g^2\, 
   \sum_{i=1}^4 \sinh^2\vp_i\, {*A_\1^i}\wedge A_\1^i- V\, {*\oneone}\,,
\label{d4trunc}
\eea

    Following analogous considerations to those in the previous two sections
we are led to make the following ansatz for 4-charge AdS bubble solutions 
of $SO(8)$ gauged four-dimensional supergravity:
\bea
ds_4^2 &=& -(H_1 H_2 H_3 H_4)^{-1/2}\, f\, dt^2 + 
(H_1 H_2 H_3 H_4)^{1/2}\, (f^{-1}\, 
   dr^2 + r^2 \, d\Omega_2^2)\,,\nn\\
A_\1^i &=& - H_i^{-1}\, dt\,,\qquad X_i= (H_1 H_2 H_3 H_4)^{1/4}\, H_i^{-1}\,,
\qquad  \cosh\vp_i= (R H_i)'\,,\nn\\
f&=& 1 +  4 g^2\, r^2\, H_1 H_2 H_3 H_4\,,\label{d4ans}
\eea
Substituting into the equations of motion following from (\ref{d4trunc}),
we find that they are indeed satisfied, provided that the functions $H_i$
obey the equations
\be
R^{-1}\, f\, (R H_i)'' 
        = - g^2\, [{(R H_i)'}^2- 1]\, (H_1 H_2 H_3 H_4)\, H_i^{-1}\,,
\ee
where in this case we have defined $R=2r$, and a prime denotes
$\del/\del R$.  These solutions in general preserve $\ft18$ of the
supersymmetry.\footnote{To be precise, the 4-charge solutions will 
preserve either $\ft18$ of the supersymmetry or none, depending on
a sign choice in the expressions for the gauge potentials.  This is
the same phenomenon as is seen in standard 4-charge BPS black holes
in four dimensions.}   In special cases where three of the $H_i$ are set to
1, the solutions will generically preserve $\ft12$ supersymmetry; if
two of the $H_i$ are set to 1, the solutions will generically preserve
$\ft14$ supersymmetry; and if one of the $H_i$ is set to 1 the
solutions will generically preserve $\ft18$ supersymmetry.

\section{$\ft18$-BPS Geometries in Six Dimensions}

     Gauged supergravity in $D=6$ \cite{romans6} follows a different 
pattern from those in other
dimensions.  The maximum number of supercharges in the $D=6$ gauged
supergravity is 16, rather than the 32 in dimensions 7, 5 and 4.  
This is because its AdS$_6$ vacuum is related to the
D4-D8 system \cite{oz}.
The pure gauged supergravity theory can be obtained from a
consistent Pauli sphere reduction, 
starting from massive type IIA
supergravity in $D=10$, and reducing on a warped 
$S^4$ hemisphere \cite{clpsix}.  As discussed in \cite{cgublp}, one can
presumably augment the theory with ${\cal N}=2$ matter.
We are interested in such a theory with $U(1)^2$ gauge fields and 
four scalars.  The consistent
reduction of such a four-scalar system, and its resulting six-dimensional
scalar potential, were obtained in \cite{cgublp}.  Following a pattern 
analogous to those for the other theories discussed
in the paper, we expect that the relevant
Lagrangian for our purpose should be given by
\bea
{\cal L}_6 &=& R\, {*\oneone} - \ft12 \sum_{i=1}^2\,
   {*d\vp_i}\wedge d\vp_i 
-\ft12 \sum_{\a=1}^2 {*d\phi_\a}\wedge d\phi_\a
- \ft12 \sum_{i=1}^2 X_i^{-2}\, {*F_\2^i}\wedge
        F_\2^i \nn\\
&&-2 g^2\,
   \sum_{i=1}^2 \sinh^2\vp_i\, {*A_\1^i}\wedge A_\1^i- V\, {*\oneone}\,,
\label{d6trunc}
\eea
where
\be
X_1=e^{-\ft1{2\sqrt2} \phi_1  -\ft1{\sqrt2}\phi_2}\,,\qquad 
X_2=e^{-\ft1{2\sqrt2} \phi_1 + \ft1{\sqrt2}\phi_2}\,.
\ee
The scalar potential is given by \cite{cgublp}
\be
V=-\ft12 g^2 \Big( (\sum_{i=1}^4 Y_i)^2 - 2 \sum_{i=1}^4 Y_i^2 +
\ft83 Y_0\sum_{i=1}^4 Y_i - \ft89 Y_0^2\Big)\,,
\ee
where
\bea
&&Y_1 = X_1\, e^{-\varphi_1}\,,\qquad
Y_2 = X_1\, e^{\varphi_1}\,,\qquad
Y_3 = X_2\, e^{-\varphi_2}\,,\qquad
Y_4 = X_2\, e^{\varphi_2}\,,\nn\\
&&Y_0\equiv (Y_1 Y_2 Y_3 Y_4)^{-3/4}= (X_1 X_2)^{-3/2}\,.
\eea
Following analogous considerations to those of the previous 
sections, and guided by
the AdS$_6$ black hole solutions obtained in $D=6$ \cite{clpsix}, we find that
the 2-charge AdS bubble solutions are given by
\bea
ds_6^2 &=& - (H_1 H_2)^{-3/4}\, f\, dt^2 +
(H_1\, H_2)^{1/4} (f^{-1}\, dr^2 + r^2\, d\Omega_4^2)
\,,\nn\\
A_i &=& -H_i^{-1}\, dt\,,\qquad 
X_i= (H_1 H_2)^{3/8}\,  H_i^{-1}\,,\qquad
\cosh\varphi_i=(R\,H_i)'\,,\nn\\
f &=& 1 + \ft49 g^2 r^2\, H_1\, H_2\,,
\eea
where $R=(2/3)^{4/9}\, r^3$, and a prime denotes a derivative with
respect to $R$.
The full set of equations of motion now reduces to
\be
R^{\ft13}\, f\, (R\,H_i)'' = - g^2 [{(R H_i)'}^2\,-1]\, 
(H_1 H_2)\, H_i^{-1}\,.
\ee
The solution preserves $\ft14$ of the supersymmetry of six-dimensional
gauged supergravity, but since this theory itself has only 16 supercharges,
the solution preserves $\ft18$ of the maximal supersymmetry
from the point of view of massive ten-dimensional type IIA supergravity.

   Having obtained the solutions in $D=6$, we can lift them back
to massive ten-dimensional type IIA supergravity, by combining
results from \cite{clpsix} and \cite{cgublp}.  The reduction ansatz
that encompasses the solutions we are considering here can be written as
\bea
ds_{10}^2 &=& \mu_0^{1/12}\,  (X_1 X_2)^{1/16}\, 
\Big(\Delta^{3/8}\, ds_6^2 +
g^{-2} \Delta^{-5/8}\,  T_{\a\beta}^{-1}
D\mu^\alpha\, D\mu^\beta \Big)\,,\nn\\
e^{\ft12\phi} {*F}_\4
&=& g\sum_{\alpha}(2Y_\alpha^2 \mu_\alpha^2 -\Delta Y_\alpha)
\epsilon_\6 -\ft13 g\Delta\, X_0\, \epsilon_\6 \nn\\
&&+\ft1{2}g^{-2}\,  T_{\a\gamma}^{-1}\, T_{\beta\delta}^{-1}\, 
  {*F_\2^{\gamma\delta}\, }\wedge D\mu^\alpha\wedge
D\mu^\beta + \ft{1}{2} g^{-1}\, T_{\a\beta}^{-1}\, {*D T^{\a\gamma}}\,  
  \wedge( \mu^\gamma\, D\mu^\beta) \,,\nn\\
e^{\phi} &=& \mu_0^{-5/16} \, \Delta^{1/4}\,  (X_1 X_2 )^{-5/8}\,,
\eea
where $D\mu^\alpha = d\mu^\alpha + g A_\1^{\alpha\beta}\, \mu^\beta$, 
$\Delta= T_{\a\beta}\, \mu^\a\, \mu^\beta$, and
\be
T_{\a\beta}\equiv \hbox{\rm diag}(Y_0,Y_1, Y_2,Y_3,Y_4)\,.
\ee
Note that the $\a$, $\beta$ indices range from 0 to 4, that $\mu^\a\,
 \mu^\a=1$, and that $A_\1^{12}=
 A_\1^1$ and $A_\1^{34}= A_\1^2$, with all other non-symmetry related 
components of $A_\1^{\a\beta}$ vanishing.

\section{Conclusions}

   The $\ft12$-BPS AdS bubble solutions obtained in \cite{linlunmal}
can be thought of as generalisations of $\ft12$-BPS AdS black hole
solutions in five or seven dimensional gauged supergravity.  In this
paper, we have extended these solutions by invoking additional gauge
fields and scalar fields in the supergravity Lagrangians, thereby 
obtaining AdS bubble generalisations of the previously-known 
multi-charge AdS black solutions of gauged supergravity.  We have also 
obtained analogous AdS bubble solutions in four-dimensional gauged
supergravity.  Our generic solutions preserve supersymmetry fractions
$\ft14$, $\ft18$ and $\ft18$ in seven, five and four dimensions respectively,
with larger fractions $2^N$ arising as special cases when some of the
additional gauge fields and scalars are set to zero.

   Using formulae for the embeddings via sphere reductions of the four
and seven dimensional gauged supergravities in $D=11$, and of the
five-dimensional gauged supergravity in type IIB supergravity, the AdS
bubble solutions that we have obtained can be lifted to M-theory or
string theory.  We also obtained AdS bubble solutions in six-dimensional 
gauged supergravity, and described their lifting to the massive type
IIA supergravity.

\section*{Acknowledgement}

    We are grateful to Clifford Johnson for discussions.}

\end{document}